# Observation of Weyl nodes in TaAs

B. Q. Lv[1,2,*], N. Xu[2,3,*], H. M. Weng[1,4,*], J. Z. Ma[1,2], P. Richard[1,4], X. C. Huang[1], L. X. Zhao[1], G. F. Chen[1,4], C. Matt[2], F. Bisti[2], V. N. Strocov[2], J. Mesot[2,3,5], Z. Fang[1,4], X. Dai[1,4], T. Qian[1,§], M. Shi[2,§], and H. Ding[1,4,§]

[1] *Beijing National Laboratory for Condensed Matter Physics and Institute of Physics, Chinese Academy of Sciences, Beijing 100190, China*
[2] *Paul Scherrer Institute, Swiss Light Source, CH-5232 Villigen PSI, Switzerland*
[3] *Institute of Condensed Matter Physics, École Polytechnique Fédérale de Lausanne, CH-1015 Lausanne, Switzerland*
[4] *Collaborative Innovation Center of Quantum Matter, Beijing, China*
[5] *Laboratory for Solid State Physics, ETH Zürich, CH-8093 Zürich, Switzerland*

\* These authors contributed equally to this work.
§ E-mail: tqian@iphy.ac.cn, ming.shi@psi.ch, dingh@iphy.ac.cn

**In 1929, H. Weyl proposed that the massless solution of the Dirac equation represents a pair of new type of particles, the so-called Weyl fermions [1]. However, their existence in particle physics remains elusive after more than eight decades. Recently, significant advances in both topological insulators and topological semimetals have provided an alternative way to realize Weyl fermions in condensed matter as an emergent phenomenon: when two non-degenerate bands in the three-dimensional momentum space cross in the vicinity of the Fermi energy (called Weyl nodes), the low-energy excitations behave exactly as Weyl fermions. Here we report the direct observation in TaAs of the long-sought-after Weyl nodes by performing bulk-sensitive soft X-ray angle-resolved photoemission spectroscopy measurements. The projected locations at the nodes on the (001) surface match well to the Fermi arcs, providing undisputable experimental evidence for the existence of Weyl fermionic quasiparticles in TaAs.**

The massless Dirac equation in the three-dimensional (3D) momentum space can be regarded as the overlap of two Weyl fermions with opposite chirality [1, 2]. The Dirac fermionic quasiparticle is stable under the protection of some crystal symmetry in topological Dirac semimetals such as $Na_3Bi$ [3] and $Cd_3As_2$ [4]. However, a separated single Weyl node is much more robust and requires no protection of crystal symmetry. An isolated Weyl node is a sink or source of gauge field of Berry curvature, like a monopole in momentum space, and the chirality corresponds to its topological charge [5-7]. Weyl nodes appear in pairs of opposite chirality in a real material due to the "No-go theorem" [8, 9]. To obtain isolated Weyl nodes, the spin degeneracy of the electronic bands has to be removed by breaking either the inversion or the time-reversal symmetry. Although non-degenerate band crossing is not rare, finding a material with only Weyl nodes near the Fermi energy ($E_F$) is a big challenge. Recently, the noncentrosymmetric and nonmagnetic transition-metal monophosphide TaAs has been predicted to be a Weyl semimetal (WSM), and twelve pairs of Weyl nodes are expected in its 3D Brillouin zone (BZ) [10, 11]. Compared with other proposals [6, 7, 12-22] in realizing a Weyl state, the TaAs family is featured by an easy sample fabrication, a non-magnetic state and no fine-tuning of the electronic states is necessary, making experimental studies of Weyl semimetals possible. Many exotic properties induced by the Weyl nodes have been predicted and observed recently, such as surface states with Fermi arcs [23, 24] and a negative magneto-resistivity [25, 26] due to the chiral anomaly [27-29]. However, crucial evidence for Weyl nodes in the bulk states has not been observed. In this paper, by using soft X-ray angle-resolved photoemission spectroscopy (ARPES), which is sensitive to the bulk states, we report the first experimental observation of Weyl nodes in TaAs.

TaAs crystallizes in a body-centred-tetragonal structure with the nonsymmorphic space group $I4_1md$. The lattice parameters are $a = b = 3.4348$ Å and $c = 11.641$ Å. The crystal structure consists of alternating Ta and As layers (Fig. 1a). The adjacent TaAs layers are rotated by 90° and shifted by $a/2$, leading to a lack of inversion symmetry. First-principles band calculations predict that TaAs is a time reversal invariant 3D WSM with twelve pairs of Weyl nodes in the 3D BZ (Fig. 1c) [10, 11]. These Weyl nodes with opposite chirality are linked to the Fermi arcs of the surface state in the projected surface BZ.

Fermi arc-like surface states have been identified in recent vacuum ultraviolet (VUV) ARPES experiments [23, 24], which are remarkably well reproduced by the band calculations for the As termination [23]. However, due to the short escape depth of the photoelectrons excited by VUV light, the bulk electronic structure was nearly invisible in the VUV ARPES experiments. To probe the bulk electronic states of TaAs, we use soft X-ray to promote valence photoelectrons to higher kinetic energy with an escape depth much longer, thus representing the bulk band structure. This bulk nature is reflected in the shallow core level spectra shown in Fig. 2a, where the two As 3$d$ surface peaks, which are strong at $hv$ = 70 eV, almost vanish at $hv$ = 440 eV. In addition, due to the Heisenberg uncertainty principle, the increase of the photoelectron escape depth sharpens the intrinsic experimental resolution in the surface-perpendicular momentum ($k_z$) [30], allowing accurate navigation in the 3D momentum space.

The bulk nature of the electronic states is further confirmed by strong $k_z$ dispersion with varying the incident photon energy (Fig. 2). The Fermi surfaces and band dispersions exhibit a modulation along the $k_z$ direction with a period of $2\pi/c'$, where $c'$ is the half of the $c$-axis lattice constant of TaAs. The experimental bands along the Γ-Z and Γ-Σ(S)-Z high-symmetry lines are in remarkably good agreement with the bulk bands from first-principles calculations with spin-orbit coupling (SOC) included, as demonstrated in Fig. 2. The Rashba splitting near $E_F$ in Fig. 2e is an indication of strong SOC in the Ta 5$d$ orbitals. The calculations show that in the absence of SOC, the band crossing near $E_F$ within the mirror-invariant planes is protected by the mirror symmetry, leading to gapless nodal rings in the mirror plane [10]. As the SOC is turned on, the nodal rings are fully gapped, generating nonzero mirror Chern numbers for the occupied states on the mirror planes and Weyl nodes slightly off the mirror planes [10].

To identify the Weyl nodes and their locations in the momentum space, we mapped out the 3D electronic structure with a number of incident photon energies. The Fermi surface map at $hv$ = 440 eV in Fig. 3b clearly shows two date-like pockets slightly off the mirror plane. The pockets originate from an "M"-shape band perpendicular to the (010) mirror plane (Fig. 3d,e), whose tops touch $E_F$ at $k_x$ = ±0.06 π/$a$. While moving along the $k_y$ direction, the two tops gradually merge together and sink below $E_F$, and eventually this band evolves into a "Λ"-shape, as illustrated in Fig.

3h. The band parallel to the mirror plane through one of the pockets exhibits a "Λ" shape with the top at $k_y = -3.53\ \pi/a$ (Fig. 3f,g). These observed band dispersions in the $k_x$-$k_y$ plane are well consistent with the calculated Weyl-cone band structure of W1.

To verify the 3D nature of the cone, we also investigate the band dispersion along the $k_z$ direction. In this case, as shown in Fig. 3i, the "M"-shape band also evolves into a "Λ"-shape. These results demonstrate that these pockets enclose the 3D Weyl nodes W1 that are located at (±0.06, ±0.47, ±0.58) in the units of ($\pi/a$, $\pi/a$, $\pi/c'$), which are pretty close to the calculated locations (±0.059, ±0.556, ±0.592) [10].

The predicted Weyl nodes near the BZ boundary (labeled as W2) are also identified in Fig. 4. As compared with the W1 nodes, the nodes of the W2 pair are very close to each other in the calculations [10]. This is consistent with our experimental results, in which we observe a nearly "Λ"-shape band rather than an "M"-shape one along the cut perpendicular to the mirror plane (Fig. 4c,d). The band also exhibits a "Λ"-shape along both the $k_y$ and $k_z$ directions with a much smaller slope along $k_z$, which is consistent with the band calculations. The excellent agreement between the experimental and calculated band dispersions along all three directions supports the existence of the Weyl nodes W2. From the measured dispersions, the Weyl nodes W2 are determined to be at (~0, ±1.05, 0), which are very close to the calculated locations (±0.01, ±1.03,0) [10].

Finally, it is crucial to verify experimentally the relationship between the bulk Weyl nodes and the surface Fermi arcs. As illustrated in Fig. 5, the projections of the experimentally determined Weyl nodes W1 onto the (001) surface BZ are exactly located on the loci of the Fermi arcs observed by our VUV ARPES, thus providing a convincing experimental evidence that Weyl nodes in certain projected surfaces are the "source" (or "drain") for the Fermi arcs, which is a hallmark of a WSM.

In summary, by using soft X-ray ARPES, we have observed the bulk band structure of Weyl semimetal TaAs, and clearly identified pairs of Weyl nodes in the momentum space. The linear band dispersion around the Weyl nodes in the 3D momentum space, which constitutes a 4D Weyl cone, can be visualized from our measurements. The positions of the Weyl nodes, as well as the anisotropic Weyl cones, are found to be consistent with our first-principles band calculations. The projections of the observed Weyl nodes W1 onto the (001) surface are connected to the observed surface Fermi arcs, thus proving evidence for the existence of Weyl fermions in TaAs.

## Methods

**Sample synthesis.** Single crystals of TaAs were grown by chemical vapor transport. A previously reacted polycrystalline TaAs sample was filled in a quartz ampoule using 2 mg/cm$^3$ of Iodine as transporting agent. After evacuating and sealing, the ampoule was kept at the growth temperature for three weeks. Large polyhedral crystals with dimensions up to 1.5 mm are obtained in a temperature field of $\Delta T =$ 1150 °C - 1000 °C. The as-grown crystals were characterized by X-ray diffraction (XRD) using a PANalytical diffractometer with Cu Kα radiation at room temperature. Single-crystal XRD was used to determine the crystal growth orientation. The average stoichiometry was determined by energy-dispersive X-ray (EDX) spectroscopy. No $I_2$ doping was detected.

**Angle-resolved photoemission spectroscopy.** Soft X-ray ARPES measurements were performed at the Advanced Resonant Spectroscopies (ADRESS) beamline at Swiss Light Source (SLS) with a SPECS analyser, and data were collected using circular-polarized light with an overall energy resolution on the order of 50-80 meV at $T = 10$ K. VUV-ARPES measurements were performed at the Surface and Interface (SIS) beamline at SLS with a Scienta R4000 analyser. Fresh surfaces for ARPES measurements were obtained by cleaving TaAs samples *in situ* in a vacuum better than $5 \times 10^{-11}$ Torr.

**Band structure calculations.** First-principles calculations were performed using the OpenMX [31] software package. We have checked that the choices of pseudo atomic orbital basis set with Ta9.0-s2p2d2f1 and As9.0-s2p2d1, the pseudo-potentials for Ta and As and the sampling of BZ (10 × 10 × 10 *k*-grid) describe the electronic structure accurately. The exchange-correlation functional within generalized gradient approximation parameterized by Perdew, Burke, and Ernzerhof has been used [32]. The optimized lattice constants $a = b = 3.4824$ Å, $c = 11.8038$ Å and atomic sites are in agreement with the experimental values.

## Acknowledgements

We acknowledge the help in plotting Fig. 1 from Hu Miao. This work was supported by the Ministry of Science and Technology of China (No. 2013CB921700, No. 2015CB921300, No. 2011CBA00108, and No. 2011CBA001000), the National Natural Science Foundation of China (No. 11474340, No. 11422428, No. 11274362, and No. 11234014), the Chinese Academy of Sciences (No. XDB07000000), the Sino-Swiss Science and Technology Cooperation (No. IZLCZ2138954), and the Swiss National Science Foundation (No. 200021-137783).

## Author contributions

H.D., T.Q. and M.S. conceived the experiments. B.Q.L., N.X. and J.Z.M. performed ARPES measurements with assistance of C.M., F.B., V.N.S. and J.M. H.M.W., Z.F. and X.D. performed *ab initio* calculations. N.X., B.Q.L., J.Z.M., T.Q. and H.D. analyzed the experimental data. N.X., B.Q.L. and H.M.W. plotted the figures. T.Q., H.D., H.M.W., M.S. and P.R. wrote the manuscript. X.C.H., L.X.Z. and G.F.C. synthesized the single crystals.

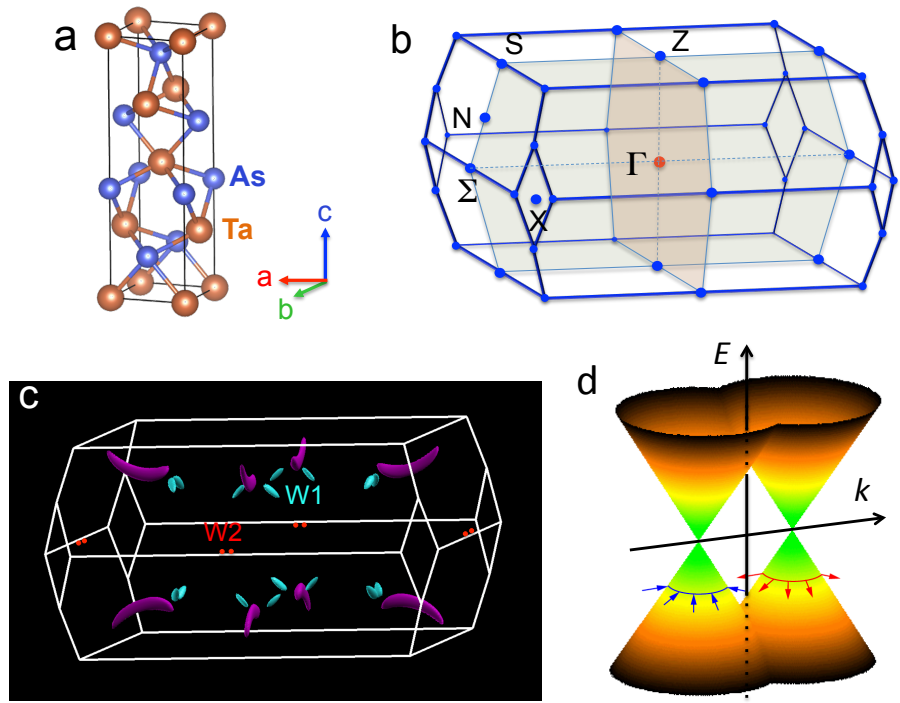

**Fig. 1. Crystal structure and electronic structure of TaAs. a**, Crystal structure of TaAs. **b**, Bulk BZ of TaAs with high-symmetry points indicated. The shaded planes represent the mirror-invariant (100) and (010) planes. **c**, Fermi surfaces from first-principles calculations. The purple banana-like hole pockets are topologically trivial Fermi surfaces without enclosing a Weyl node. The blue date-like electron pockets enclose the Weyl nodes W1, which are 21 meV below $E_F$. The Weyl nodes W2 in the $k_z = 0$ plane are only 2 meV above $E_F$, and the Fermi surfaces enclosing them are tiny and marked by the red dots. **d**, Schematic band dispersions of one pair of Weyl nodes in a 2D plane of the momentum space. The arrows indicate that the Weyl node is the monopole of Berry flux in the momentum space.

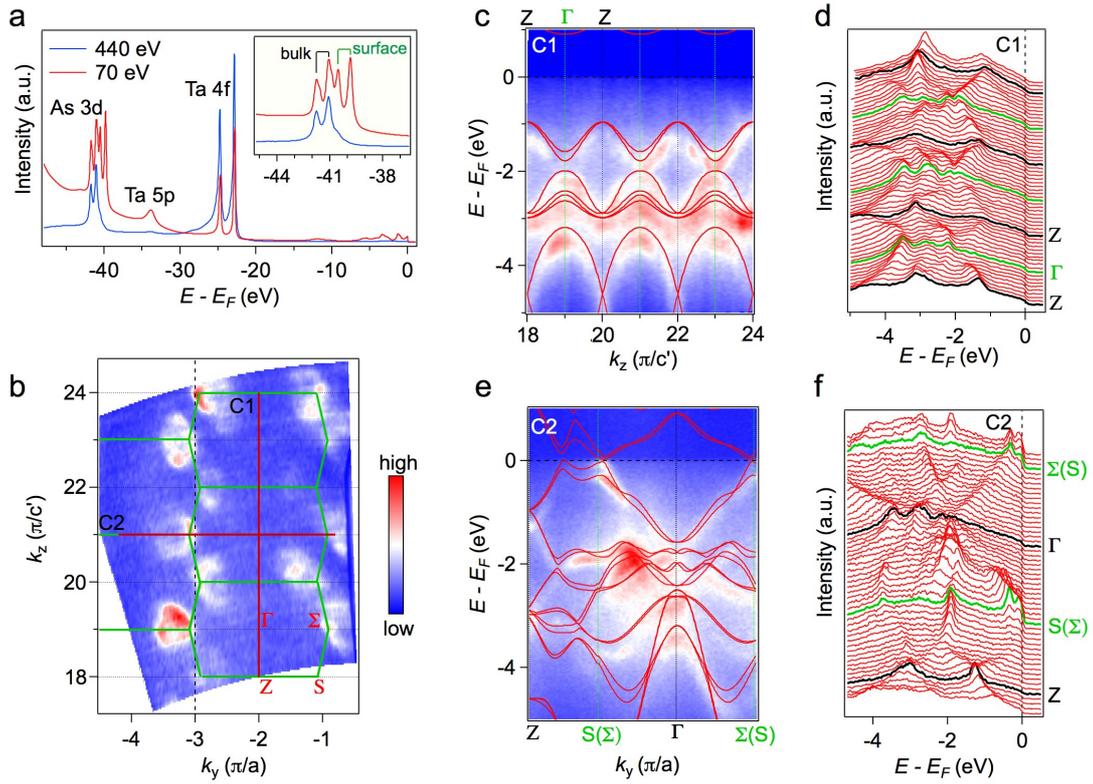

**Fig. 2. Core level spectra and electronic structure in the $k_x = 0$ plane. a**, Core level spectra taken at $h\nu = 70$ and $440$ eV, respectively. The inset shows the magnification of As $3d$ core level peaks, indicating that the surface electronic states are dramatically suppressed with the incident soft X-ray. **b**, Photoemission intensity plot at $E_F$ in the $k_y$-$k_z$ plane at $k_x = 0$. The green lines represent the BZ structure in the $k_y$-$k_z$ plane. **c and d**, Photoemission intensity plot and energy distribution curves along Γ-Z (C1) indicated in **b**, respectively. **e and f**, Photoemission intensity plot and energy distribution curves along Γ-Σ(S)-Z (C2) indicated in **b**, respectively. For comparison, the calculated bands along Γ-Z and Γ-Σ(S)-Z are plotted on top of the experimental data in **c** and **e**, respectively.

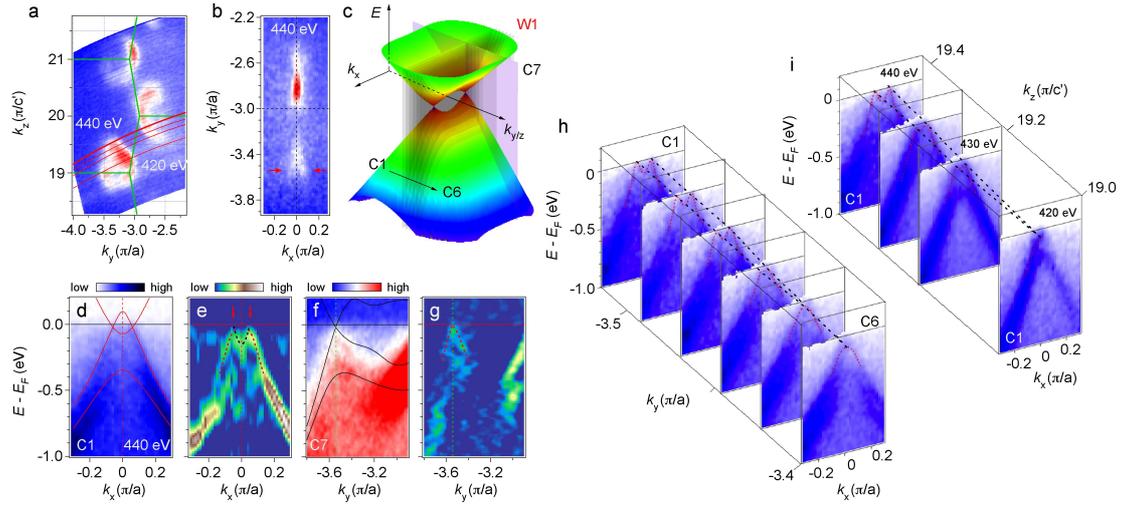

**Fig. 3. Band dispersions near the Weyl nodes W1. a**, Photoemission intensity plot at $E_F$ in the $k_y$-$k_z$ plane at $k_x = 0$. The red curves from above to below indicate the momentum locations taken at $hv$ = 440, 435, 430, and 420 eV, respectively. **b**, Photoemission intensity plot at $E_F$ taken at $hv$ = 440 eV. The arrows indicate the date-like pockets that enclose the Weyl nodes W1. **c**, Schematic band dispersion of one pair of Weyl nodes in the $k_x$-$k_y$ plane. The vertical planes C1-C6 schematize the ARPES measurements that slice through the Weyl cone at different $k_y$ positions. **d and e**, Photoemission and curvature intensity plots along the $k_x$ direction through the two Weyl nodes W1, respectively, illustrating the "M"-shape dispersion of a pair of Weyl nodes. **f and g**, Photoemission and curvature intensity plots along the $k_y$ direction through one of the Weyl nodes W1, respectively. For comparison, the calculated bands through the Weyl nodes W1 along the $k_x$ and $k_y$ directions are plotted on top of the experimental data in **d** and **f**, respectively. Note that the calculated bands in **d** and **f** are shifted by 0.11 $\pi/a$ along the $k_y$ direction to be consistent with the experimental bands. **h**, Band dispersions along the $k_x$ direction at different $k_y$ positions. **i**, Band dispersions along the $k_x$ direction at different $k_z$ positions. The dotted lines in **h** and **i**, which are guides to the eye that indicate the band dispersions, show the evolution of the band shape. The dashed lines in **h** and **i** track the band tops.

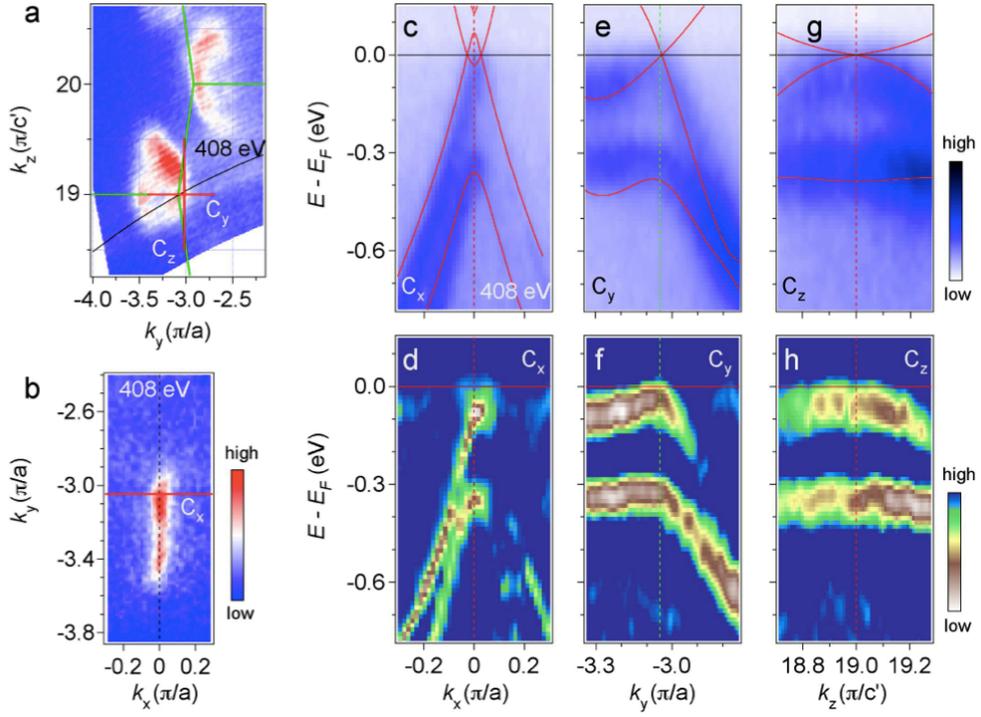

**Fig. 4. Band dispersions near the Weyl nodes W2. a**, Photoemission intensity plot at $E_F$ in the $k_y$-$k_z$ plane at $k_x = 0$. The black curve indicates the momentum location for the data recorded at $hv = 408$ eV. **b**, Photoemission intensity plot at $E_F$ taken at $hv = 408$ eV. **c and d**, Photoemission and curvature intensity plots along $C_x$ indicated in **b**, respectively. **e and f**, Same as **c** and **d** but recorded along $C_y$ indicated in **a**. **g and h**, Same as **c** and **d** but measured along $C_z$ indicated in **a**. For comparison, the calculated bands along the $k_x$, $k_y$ and $k_z$ directions through the Weyl nodes W2 are plotted on top of the experimental data in **c**, **e** and **g**, respectively.

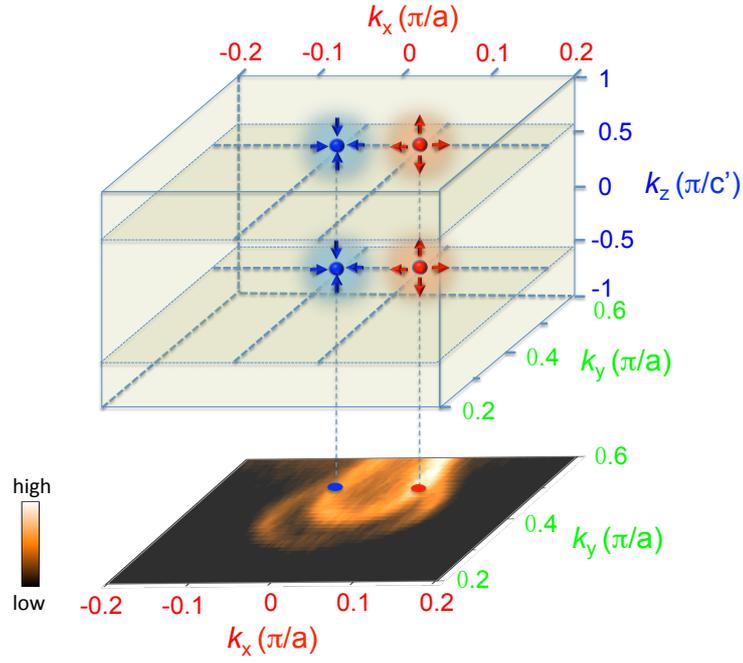

**Fig. 5. Locations of experimentally determined Weyl nodes W1 in the 3D Brillouin zone and their projections onto the measured (001) surface Fermi arcs.** The projections of the Weyl nodes are connected to the Fermi arcs, thus proving the existence of Weyl fermions in TaAs. The Weyl nodes are located at (±0.06, 0.47, 0.58) in the units of ($\pi/a$, $\pi/a$, $\pi/c'$). Red and blue colors represent the Weyl nodes with opposite chirality. The arrows indicate that the Weyl node is the monopole of Berry flux in the momentum space. The Fermi arcs are obtained at $hv$ = 54 eV using circular-polarized light.